\documentclass[a4paper]{article}

\usepackage[pages=all, color=black, position={current page.south}, placement=bottom, scale=1, opacity=1, vshift=5mm]{background}
\SetBgContents{
	\tt    
}      

\usepackage[margin=1in]{geometry} 

\usepackage{amsmath}
\usepackage{amsthm}
\usepackage{amssymb}
\usepackage{subfig}
\usepackage{graphicx}

\usepackage[utf8]{inputenc}
\usepackage{hyperref}
\hypersetup{
	unicode,
	pdfauthor={Author One, Author Two, Author Three},
	pdftitle={A simple article template},
	pdfsubject={A simple article template},
	pdfkeywords={article, template, simple},
	pdfproducer={LaTeX},
	pdfcreator={pdflatex}
}

\newcommand{\be}{\begin{equation}}
\newcommand{\ee}{\end{equation}}
\newcommand{\bea}{\begin{eqnarray}}
\newcommand{\eea}{\end{eqnarray}}

\usepackage[sort&compress,numbers,square]{natbib}
\theoremstyle{plain}

\theoremstyle{definition}

\usepackage{graphicx, color}
\graphicspath{{fig/}}

\usepackage{algorithm, algpseudocode} 
\usepackage{mathrsfs} 

\usepackage{lipsum} 

\begin{document}
\title{Conformal structure of singularities in some varying fundamental constants bimetric cosmologies}

\author{Konrad Marosek${^1} {^{\divideontimes}}$ \and Adam Balcerzak$^{2,3}$}

\date{
\small{$^1$Institute of Mathematics, Physics and Chemistry, Maritime University of Szczecin, Wa{\l }y Chrobrego 1-2, 70-500 Szczecin, Poland \\
$^2$Institute of Physics, University of Szczecin,  Wielkopolska 15, 70-451 Szczecin,  Poland \\
$^3$Copernicus Center for Interdisciplinary Studies, Szczepa\'nska 1/5, 31-011 Krak\'ow, Poland \\
$^{\divideontimes}$Corespondence: k.marosek@pm.szczecin.pl}}

\maketitle

\abstract{
In this paper, we explore the conformal structure of singularities arising from varying fundamental constants using the method of Penrose diagrams. We employ a specific type of bimetric model featuring two different metrics. One metric describes the causal structure for matter, while the other characterizes the causal structure for gravitational interactions, which is related to variations in fundamental constants such as the gravitational constant and the speed of light. For this reason, we focused on the gravitational metric to calculate the conformal transformation and compose Penrose diagrams for the singularities arising from the varying fundamental constants. We have shown that, in one case, the parameter such as the scale factor, the density and the pressure resemble those of the finite scale factor singularity (FSF).  Despite singularity appears in constant conformal time in our case and in the case of FSF the Misner-Sharp horizon looks different. Our another case is similar to sudden future singularity (SFS), but there are differences in the conformal structures. 
We have also shown that in our cases the behavior of Misner-Sharp horizon strongly depends on initial conditions. The last analytical solution which we introduced is identical to conformal structure of the standard exotic singularity for the matter.}

\section{Introduction}
\label{sec:intro}

Singularities are one of the most crucial problems in general relativity. They are typically defined in terms of the geodesic incompleteness \cite{Hawking}. The emergence of such  singularities in cosmological scenarios is usually preceded (or followed) by an extreme state of the universe in which physical quantities approach infinity. The most common example of geodesically incomplete singularity is the Big Bang, whose appearance in all of the cosmological models based on classical theory of gravity indicates their inherent incompleteness despite the evolution of the universe from the beginning of the inflation era to the present moment seems to be well understood. Some attempts which have been made to include the initial singularity into the cosmological framework by introducing pre-big-bang eras were based on the tree-level low-energy effective action of string theory \cite{Khoury,Khoury2,Steinhardt,Steinhardt2,Gasperini}.  On the other hand, the discovery of the accelerated expansion of the Universe \cite{Tonry} associated with the presence of dark energy inspired the unveiling of cosmological models incorporating exotic singularities in future evolution. Within a multitude of scenarios, each featuring an exotic singularity, distinctions in the strength of these singularities are outlined by the Tipler and Kr\'olak  criteria \cite{TiplerDef,KrolakDef}. These variations include the type I big-rip singularity \cite{Caldwell,Dabrowski}, the sudden future singularity (SFS or type II) \cite{Barrow}, finite scale factor singularities (FSF or type III) \cite{Odintsov1,FSF}, big-separation singularities (BS or type IV) \cite{Odintsov2}, and the $w$-singularity (type V) \cite{w}, all occurring within finite time. Additionally, there are the little-rip and pseudo-rip singularities that emerge as time extends to infinity \cite{Frampton,Frampton1} (for a comprehensive description of various types of cosmological singularities, see \cite{Haro}). Another criterion for differentiating the previously mentioned exotic singularities is their adherence to or violation of certain energy conditions (null, weak, strong, dominant) due to the presence of dark energy and phantoms  \cite{Dabrowski}.

The early concept of varying physical constants, initially considered by Weyl \cite{Weyl} and Eddington \cite{Eddington1,Eddington2}, was further developed by Dirac through his Large Numbers Hypothesis \cite{Dirac}. Dirac identified a numerical relationship between cosmological and microscopic scales, inspiring the notion that the gravitational constant could vary being inversely proportional to time. This idea, in turn, influenced the creation of the Brans-Dicke scalar-tensor gravity model \cite{BransDicke}, wherein the gravitational constant becomes inversely proportional to the scalar field.  Afterward, models were developed wherein various physical constants could undergo variations, including the speed of light $c$ \cite{Barrow2,AlbrechtMagueijo,BarrowMagueijo}, electron charge $e$, the proton-to-electron mass ratio $m_p / m_e$, or the fine-structure constant $\alpha$ \cite{Beckenstein1,Beckenstein2}. Notably, some physical constants are interconnected, and the variation of one may imply changes in others. Theories involving varying constants offer solutions to standard cosmological problems such as the flatness problem, the horizon problem, the $\Lambda$-problem \cite{Barrow2,AlbrechtMagueijo}, or the singularity problem \cite{Marosek1}.

In this paper, we explore the conformal structure of specific spacetimes that serve as solutions within a certain bimetric gravity model introduced in \cite{Marosek&Balcerzak}. In this variant, both the effective gravitational constant and the effective speed of light undergo time variation. An intriguing aspect of this model is that having a non-singular metric defining the causal structure for matter, while simultaneously featuring a singular metric for the gravitational sector in solutions, does not necessarily compromise the consistency of the model. This insight suggests specific features in the conformal structure of the various types of singularities induced by the variation of the fundamental constants $c$ and $G$ in the metric defining the causal structure for gravitational interactions.

The paper is structured as follows. In Section \ref{sec2}, we present an overview of the bimetric model incorporating the varying fundamental constants $c$ and $G$. Section \ref{sec3} discusses the conformal mapping of the spacetime whose geometry is given by the gravitational metric onto the Einstein's static universe. The analysis of the conformal structure of the singularities resulting from variations in fundamental constants is covered in Section \ref{sec4}. Finally, Section \ref{sec5} provides a summary of our findings and concluding remarks.

\section{A bimetric model with varying fundamental constants}
\label{sec2}

In this paper, we employ a bimetric model incorporating two metrics: one, denoted as $g_{\mu \nu}$, pertains to matter, while the other, represented by ${\hat{g}}_{\mu \nu}$, characterizes the causal structure for the gravitational field \cite{Marosek&Balcerzak}. Let us notice that the matter metric exhibits self-similarity, as it admits a homothetic vector field \cite{Maartens}. The following relation between these metrics:
\be
\label{MetricForm1}
{\hat{g}}_{\mu \nu}= g_{\mu \nu} {\left[ \alpha - \left( \alpha - 1 \right) \left( {\delta}_{ 0 \mu} {\delta}_{ 0 \nu} \right) \right]}^2 ~,
\ee
where $ \alpha = \alpha \left( t \right) $ is a dimensionless time dependent function,  aligns with the framework of disformal gravity \cite{Bekenstein}, a distinct category within bimetric theories. 
The time components are the same in both metrics $ {\hat{g}}_{0 0}=g_{0 0}~$, while spatial parts of gravitational metric are scaled by the function $ {\alpha}^2 $  in comparison to the matter metrics:
\be \label{g11}
{\hat{g}}_{1 1}= {\alpha}^2 g_{1 1}~,
\ee
\be \label{g22}
{\hat{g}}_{2 2}= {\alpha}^2 g_{2 2}~,
\ee
\be \label{g33}
{\hat{g}}_{3 3}= {\alpha}^2 g_{3 3}~.
\ee
In such theory, light follows the paths dictated by the matter metric, which means the speed of light might change when observed from the viewpoint of spacetime determined by the gravitational metric. 
The connection between the two metric also suggests that the cosmological expansion could affect the transmission of gravitational waves differently than it does for light (the time delay between the arrival of gravitational waves and light signals has been measured in binary neutron star systems \cite{Ligo,Ligo2}).

The relationship (\ref{MetricForm1}) emerges naturally in the context of disformal gravity \cite{Bekenstein} a specialized form of bimetric gravity theory that is based on the most general mapping between two metrics, incorporating a single scalar field while preserving spacetime diffeomorphisms. Typically, this mapping is represented by:
\be
\label{DisformalMetric}
{\tilde{g}}_{\mu \nu}= C \left( \phi , X \right) g_{\mu \nu} + D \left( \phi, X \right) {\partial}_{\mu} \phi {\partial}_{\nu} \phi ~,
\ee
where $X = {\partial}_{\mu} \phi {\partial}^{\mu} \phi$ and $\phi$ represents a kinetic term for the scalar field $\phi$, and  $C \left( \phi , X \right)$ and $D \left( \phi, X \right)$ are general functions dependent on $\phi$ and $X$. To demonstrate that (\ref{MetricForm1}) can be derived from (\ref{DisformalMetric}) we select $C \left( \phi , X \right)\equiv\phi^2$ and $D \left( \phi, X \right) = f \left( \phi \right) + g \left( \phi \right) {\partial}_{\mu} \phi {\partial}^{\mu} \phi$ such that $f(\phi)$ and $g(\phi)$ are some yet-to-be-defined functions of the scalar field $\phi$. Adopting a homogeneous and isotropic cosmological framework (achieved by assuming the Friedmann-Lemaître-Robertson-Walker line element) requires the scalar field to be spatially constant, which in an expanding universe implies $\phi=\phi(t)$. These two assumptions lead us to a specific relationship between the diagonal components (the only non-zero components in this model) of the two metrics:
\bea
\label{relg00}
{\tilde{g}}_{0 0} &=& {\phi}^2 g_{0 0} + f \left( \phi \right) {\dot{\phi}}^2 + g \left( \phi \right) {\dot{\phi}}^4 g^{0 0}~, \\
\label{relgkk}
{\tilde{g}}_{k k} &=& { \phi}^2 g_{k k}~.
\eea
Equation (\ref{relgkk}) already establishes the relationship between the spatial components of the two metrics as given by Eqs. (\ref{g11})-(\ref{g33}). Assuming $g_{00}=g^{00}=-1$ (achievable by choosing the time coordinate $x^0$ as the proper time of comoving observers) results in the relationship between the temporal components of the two metrics, as expressed by $ {\hat{g}}_{0 0}=g_{0 0}~$, provided that
\be \label{govphi}
-{\phi}^2 + f \left( \phi \right) { \dot{\phi} }^2 - g \left( \phi \right) { \dot{\phi} }^4 = -1~.
\ee

The total action in the considered model reads:
\be
\label{Action}
S= S_g[\hat{g}] + S_{matter}[g]~.
\ee
where the gravitational action 
\be
\label{gravaction}
S_g[\hat{g}]= -\frac{1}{16 \pi G_0} \int d^4 x R \left[ \hat{g} \right] \sqrt{-\hat{g}} ~.
\ee
represents the Einstein-Hilbert action calculated with the gravitational metric $ \hat{g}_{\mu\nu}$, while  the matter action given by
\be
\label{fieldaction}
S_{matter}[g]=-\frac{1}{2 {c_0}} \int d^4 x L_{matter} \sqrt{-g}~,
\ee
depends on the matter metric $g_{\mu\nu}$ only. The quantities $G_0$ and $c_0$ in the formula above are some constants of the same dimensions as the dimensions of the Newton constant $G$ and the speed of light $c$, respectively.

By varying the action (\ref{Action}) respect to the $g^{ \mu \nu }$ we obtain the field equations with the following  temporal and spatial components:
\bea
\label{eom00}
{\alpha}^{3} \left(R_{0 0} \left[\hat{g}\right] - \frac{1}{2} {\hat{g}}_{0 0} R \left[\hat{g}\right] \right) = \frac{8 \pi G_0}{{c_0}^4} T_{0 0}~,\\
\label{eom11}
\alpha \left(R_{i i} \left[\hat{g}\right] - \frac{1}{2} {\hat{g}}_{i i} R \left[\hat{g}\right] \right) = \frac{8 \pi G_0}{{c_0}^4} T_{i i}~.
\eea

Some remarks should be made about the variational principle discussed above. In our model, the function $\alpha$ is explicitly a function of time. This time dependence can arise, for instance, from (\ref{govphi})  if the model is seen as emerging from disformal gravity theory, where we identify $\phi\equiv\alpha$. With a specific ansatz for the time dependence of $\alpha(t)=\phi(t)$,  Eq. (\ref{govphi}) can still be satisfied at least locally (near the singularity), provided that $f(\phi)$ and $g(\phi)$ are appropriately chosen and satisfy the condition given by (\ref{govphi}). Therefore, varying the action (\ref{Action}) with respect to the metric components $g_{\mu\nu}$ alone should be sufficient to derive the complete set of equations governing the evolution of the cosmological framework.

For the Friedmann metric corresponding to the matter specified by:
\be
\label{FriedmanMetric}
ds_{M}^2= -{c_0}^2 dt^2 + a^2 (t) \left[ \frac{dr^2}{1-kr^2} + r^2(d{\theta}^2+ {\sin}^2 \theta d {\phi}^2) \right]~,
\ee
where $k=0, \pm1$ is the curvature index, the gravitational metric takes the following form:
\be
\label{NewMetric}
ds_{G}^2= -{c_0}^2 dt^2 + {\alpha}^2 a^2 (t) \left[ \frac{dr^2}{1-kr^2} + r^2(d{\theta}^2+ {\sin}^2 \theta d {\phi}^2) \right]~.
\ee
Inserting (\ref{FriedmanMetric}) and (\ref{NewMetric}) into the field equations (\ref{eom00}) and (\ref{eom11}), while assuming $k=0$, yields the equations describing the density $ \rho \left( t \right) $ and the pressure $ p \left( t \right) $:
\bea
\label{Density}
\rho \left( t \right) = \frac{3 {\alpha}^3 \left( t \right) }{8 \pi G_0} \left( \frac{{ \dot{a}}^2 \left( t \right) }{a^2 \left( t \right) } + \frac{2 \dot{a} \left( t\right) \dot{\alpha} \left( t \right)}{ a \left( t \right) \alpha \left( t \right)} + \frac{ { \dot{\alpha}}^2 \left( t \right) }{{\alpha}^2 \left( t\right) } \right)~,
\eea
\bea
\label{Pressure}
p \left( t \right) = - \frac{ {c_0}^2 \alpha \left( t \right)}{8 \pi G_0} \left( \frac{{ \dot{a}}^2 \left( t \right) }{a^2 \left( t \right) } + \frac{6 \dot{a} \left( t\right) \dot{\alpha} \left( t \right)}{ a \left( t \right) \alpha \left( t \right)} \right. + \left. \frac{ { \dot{\alpha}}^2 \left( t \right) }{{\alpha}^2 \left( t\right) }   +  \frac{2 \ddot{a} \left( t \right)}{ a \left( t \right)} + \frac{2 \ddot{\alpha} \left( t \right)}{ \alpha \left( t \right)} \right)~.
\eea
The continuity equation is given by
\be
\label{ConEqu}
\dot{\rho} \left( t \right) + 3 \frac{ \dot{a} \left( t \right) }{ a \left( t \right) } \left( \rho \left( t \right) + \frac{{ \alpha}^2 \left( t \right)}{{c_0}^2}p \left( t \right) \right) + 3 \frac{\dot{\alpha} \left( t \right) }{\alpha \left( t \right)} \left( \frac{{\alpha}^2 \left( t \right)}{{c_0}^2} p \left( t \right) \right) = 0 ~.
\ee
To maintain consistency with the disformal gravity model, $\alpha$ needs to be equated to the scalar field. The equations (\ref{Density}) and (\ref{Pressure}) are similar to the field equations in Brans-Dicke model when $ \alpha \equiv \phi $ which yields that $ \alpha $ is linked to the variation of $G$ and $c$. In this model the gravitational constant and the speed of light are dynamical parameters which can be expressed as follows:
\bea
\label{gvarying}
G \left( t \right) = \frac{G_0}{{\alpha}^3 \left( t \right)}~, \\
\label{cvarying}
c \left( t \right) = \frac{c_0}{\alpha \left( t \right)}~.
\eea
The assumption of $\alpha = const.$ change the equations (\ref{Density}), (\ref{Pressure}) and (\ref{ConEqu}) into the standard Friedmann equations where the proper value of gravitational constant and the speed of light are given, respectively, by the following formulas:
\bea
G_{FM} = G_0 / {\alpha}^3~,
\eea
\bea
c_{FM} = c_0 / \alpha~.
\eea

\section{Conformal mapping of the gravitational spacetime onto Einstein's static universe}
\label{sec3}
The portion of the complete solution for the bimetric model introduced in Section \ref{sec2}, related to the gravitational sector as given by (\ref{NewMetric}), may exhibit singular behavior due to variations in the fundamental constants $c$ and $G$ (see eqs. (\ref{gvarying}) and (\ref{cvarying})). The structure of singular cosmological spacetimes, including their causal relationships and the nature of singularities, is most effectively illustrated using Penrose diagrams \cite{Hawking}. These diagrams emerge through the process of compactification, achieved through specific conformal mappings of the particular spacetimes onto Einstein's static universe. The conformal time  $\eta$ for the flat variant ($k=0$) of the gravitational metric (\ref{NewMetric}) is as follows:
\bea
\label{conftime}
\eta = \int \frac{c_0 dt}{a \left( t \right) \alpha \left( t \right)}~.
\eea
The gravitational metric (\ref{NewMetric}) expressed in terms of the new variable $\eta$ is:
\bea
\label{Metric1}
ds^2_{G}  = d \hat{s}^2 a^2 \left( \eta \right) {\alpha}^2 \left( \eta \right)~,
\eea
where
\bea
\label{MinkowskiMetric}
d \hat{s}^2 = - d {\eta}^2 + dr^2 + r^2 \left( d{\theta}^2+ {\sin}^2 \theta d {\phi}^2 \right)~,
\eea
is Minkowski metric. With the following coordinate transformations:
\bea
t' = \arctan \left( \eta + r \right) + \arctan \left( \eta - r \right)~,
\eea
\bea
r' = \arctan \left( \eta + r \right) - \arctan \left( \eta - r \right)~,
\eea
where $ 0 \leq r \leq \infty $, the spacetime given by the line element (\ref{MinkowskiMetric}) can be mapped onto the Einstein static universe described by the following metric \cite{Hawking}:
\bea
d s^2_E = - d {t'}^2 + d{r'}^2 + {\sin}^2 r' \left( d{\theta}^2+ {\sin}^2 \theta d {\phi}^2 \right)~.
\eea
Let us note that the conformal time of the gravitational metric differs from that of the standard Friedman metric (with non-scaled spatial part), although the derivation of the formula is analogous.

\section{Conformal structure of singularities induced by variation of the fundamental constants}
\label{sec4}

It has been shown that in general relativity and various theories of gravitation, singularities can appear in finite time. This can be caused by the presence of matter fields, such as phantoms or ghosts, which violate energy conditions. In the standard Friedman model, the Sudden Future Singularity can occur when the Hubble expansion rate $H_s=\dot{a}/a < \infty$ and the acceleration $ \ddot{a}/a \rightarrow - \infty$ reach specific values at the singularity time $t_s$, which is clearly related to the density $\rho_s < \infty$ and pressure $p_s \rightarrow \infty$.  It should be noted that in this case, only the dominant energy condition is violated. This heuristic reasoning led to the following form of the scale factor describing the evolution of the universe, which ends in a singularity \cite{Barrow}:

\bea
\label{ScaleFactor1}
a_B \left( t \right) = a_s \left[ \delta +\left( 1 + \delta \right) {\left( \frac{t}{t_s} \right)}^m  - \delta {\left( 1 - \frac{t}{t_s} \right)}^n \right],
\eea 
where $ \delta $, $ a_s $, $ t_s $, $ n $, $ m $ are some constants. Although (\ref{ScaleFactor1}) has been received in the case of SFS singularity then can also describe other types of future singularities. There are also different scale factor to the (\ref{ScaleFactor1}) that can simulate future singularities for instance \cite{Marosek1}:
\bea
\label{ScaleFactor2}
a_M \left( t \right) = a_s { \left( \frac{t}{t_s} \right)}^m \exp {\left( 1 - \frac{t}{t_s} \right)}^n~.
\eea
The same singular behaviors arise as $t \rightarrow t_s$ for both scale factors (\ref{ScaleFactor1}) and (\ref{ScaleFactor2}). The finite scale factor singularity (FSF) emerges for $0 < n < 1$ ($a(t) \rightarrow a_s$, $\rho \rightarrow \infty$, $|p| \rightarrow \infty$). For $1 < n < 2$, one obtains the sudden future singularity (SFS) ($a(t) \rightarrow a_s$, $\rho \rightarrow \rho_s$, $|p| \rightarrow \infty$). For $n > 2$, the $w$-singularity occurs ($a(t) \rightarrow a_s$, $\rho \rightarrow 0$, $|p| \rightarrow 0$, $w$-index $\rightarrow \infty$) \cite{w}.

In order to investigate the singularity associated with varying constants, we assume the following scale factor for the matter metric:
\bea
\label{ScaleFactor3}
a \left( t \right) = a_0 t^m~,
\eea
where $ a_0 $ and $ m $ are constants. The scale factor (\ref{ScaleFactor3}) exclusively governs the initial singularity and does not directly influence future singularities. In contrast, the scalar field $\alpha$ in the gravitational metric (\ref{NewMetric}) demands an opposing behavior, as expressed by:
\bea
\label{ScaleFactor4}
a \left( t \right) \alpha \left( t \right) = a_0 t^m {\left( 1 - \frac{t}{t_s} \right)}^n~,
\eea
where $ n $ is constant. 

It is possible to find the analytical solution for the conformal time (\ref{conftime}) for certain values of $m$ and $n$ parameters in the gravitational scale factor (\ref{ScaleFactor4}). One of the solutions is obtained for $m=1/2$ and $n=1/2$. The conformal time expressed in this case is:
\bea
\label{ConfTimeS1}
{\eta} = \frac{2 \sqrt{t_s}}{a_0} \arcsin \left[ \sqrt{\frac{t}{t_s}} \right]~.
\eea
In more detail, for $0 \leq t \leq t_s$ we have:
\bea
\label{ConfTimeRangeS1}
0 \leq {\eta} \leq \frac{ \pi \sqrt{t_s}}{a_0}~.
\eea
In such a case, as $t \rightarrow 0$, the scale factor $a \rightarrow 0$, and the density for both the matter and gravitational metric $\rho \rightarrow \infty$, along with the pressure (as given by  (\ref{Pressure})) $p \rightarrow \infty$, which corresponds to the Big Bang singularity. At the time $t_s$, the scale factor $a \rightarrow a_0 \sqrt{t_s}$, the scalar field $\alpha \rightarrow 0$, and both the density and pressure for the gravitational metric become singular ($\rho \rightarrow \infty$, $p \rightarrow \infty$). In terms of the behavior of the scale factor, the density, and the pressure, this type of singularity is similar to the finite scale factor singularity (FSF) in the matter sector.

One is able to invert the relation (\ref{ConfTimeS1}) to obtain:
\bea
\label{InvertConfTimeS1}
t \left( \eta \right) = t_s {\sin}^2 \left[ \frac{ a_0 \eta}{ 2 \sqrt{t_s}} \right]~.
\eea
Analyzing the conformal time $\eta$, it can be observed that the model starts with the Big Bang singularity at hypersurface $\eta = 0$  ($t = 0$) and extends to hypersurface $ \eta = \pi \sqrt{ t_s } / a_0 $, which is spacelike.

Using the definition of the affine parameter
\bea
\label{AffineParameter}
\lambda \left( t \right) = \int a(t) \alpha(t) dt~,
\eea
we get:
\bea
\label{AffineParameterS1}
\lambda \left( t \right) = \frac{a_0 t_s}{4} \left[\sqrt{t- \frac{t^2}{t_s}} \left( \frac{2t}{t_s} -1 \right) \right. + \left. \sqrt{t_s} \arcsin \left[ \sqrt{\frac{t}{t_s}} \right] \right]~,
\eea 
so that at the Big Bang singularity, $ \lambda(0) = 0 $, and $ \lambda ( t_s ) = (1/8) a_0 \pi {t_s}^{3/2} $, indicating that the affine parameter remains finite.

In the case of the considered model the calculation of Misner-Sharp $M$ mass \cite{MisnerSharpMass,Harada} gives:
\bea
\label{SharpMisnerMassS1}
\frac{2 M}{a \left( t \right) \alpha \left( t \right)} = \frac{ {a_0}^2 r^2 \left( t_s - t \right) {\left( t_s - 2t \right)}^2}{ 4 {t_s}^3 \sqrt{ t^2 - \frac{t^3}{t_s}}}~,
\eea
and thus, the past/future trapping regions occur for
\bea
\label{SharpMisnerHorizon}
r &>& \frac{ 2 {t_s}^\frac{3}{2} {\left[ { t^2 \left( \eta \right)  - \frac{t^3\left( \eta \right)}{t_s} } \right]}^{\frac{1}{4}}}{ a_0 \left[ t_s - 2t \left( \eta \right) \right] \sqrt{ t_s - t \left( \eta \right)}}~.
\eea

\begin{figure}

  \begin{minipage}[b]{0.37\textwidth}
    \includegraphics[width=\linewidth]{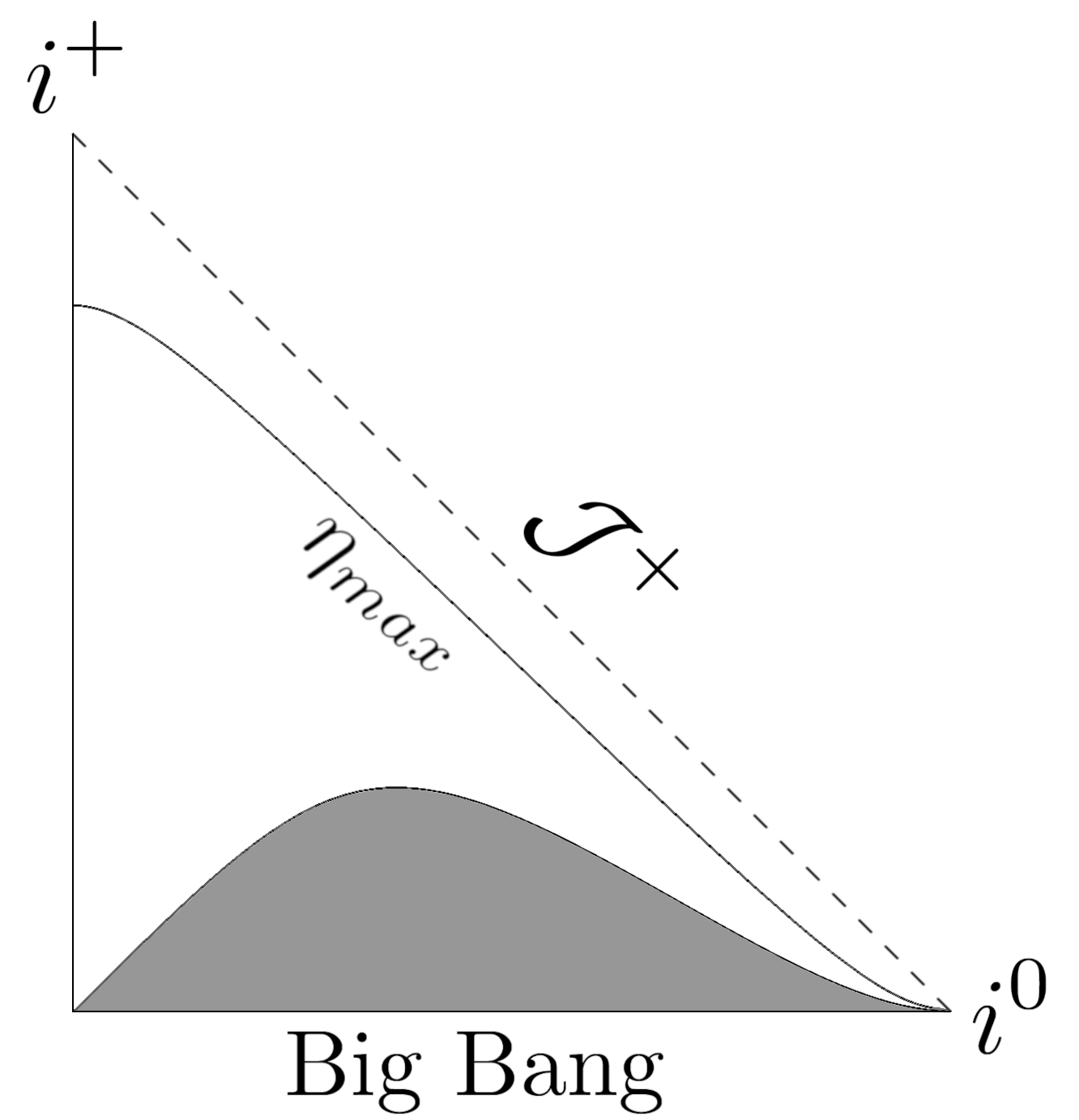}
    \captionsetup{width=1.3\linewidth} 
    \caption{Penrose diagram for the model given by  (\ref{ScaleFactor4}) for $ m = 1/2 $, $ n = 1/2 $, $ a_0=1$, $ t_s = 1$ with $ \eta_{max}= \pi $.}
    \label{plot1}
  \end{minipage}
  \hspace{0.1\textwidth}
  \begin{minipage}[b]{0.37\textwidth}
    \includegraphics[width=\linewidth]{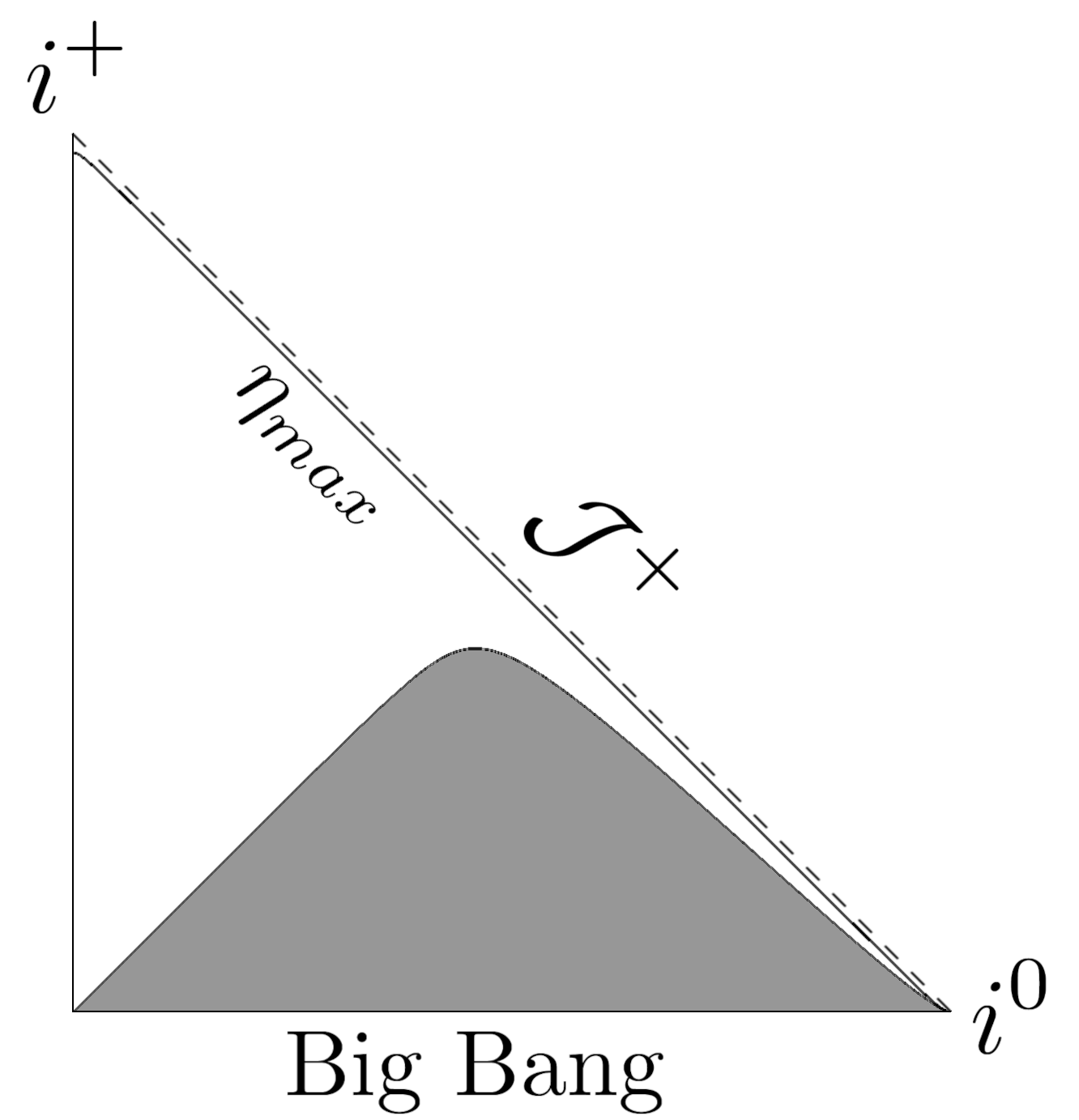}
     \captionsetup{width=1.3\linewidth}
    \caption{Penrose diagram for the model given by  (\ref{ScaleFactor4})  for $ m = 1/2 $, $ n = 1/2 $, $ a_0=1$, $ t_s = 100 $ with $ \eta_{max}= 10 \pi $.}
    \label{plot2}
  \end{minipage}
  \hspace{0.1\textwidth}
\centering
  \begin{minipage}[b]{0.37\textwidth}
    \includegraphics[width=\linewidth]{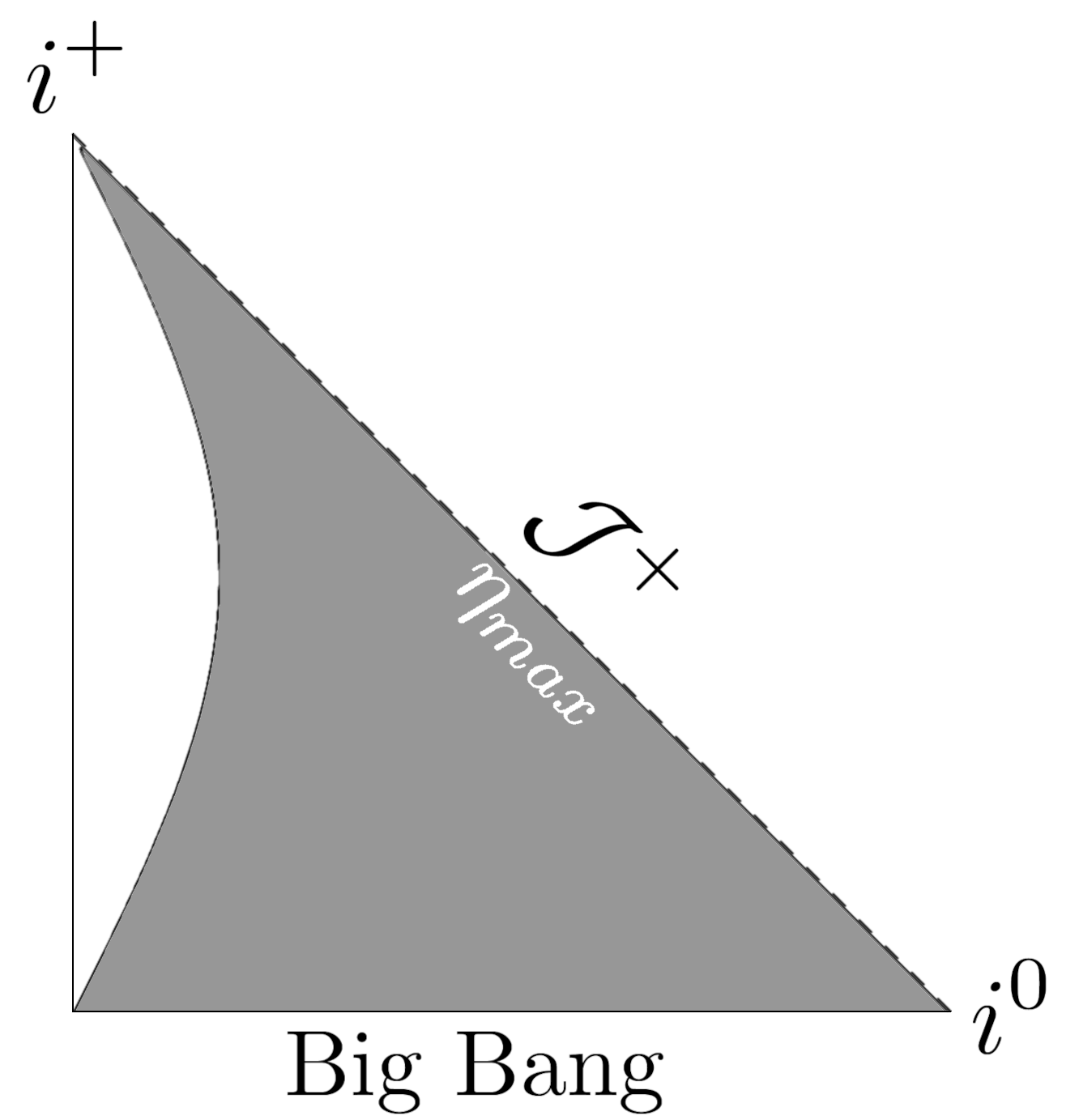}
     \captionsetup{width=1.3\linewidth}
    \caption{Penrose diagram for the model given by  (\ref{ScaleFactor4})  for $ m = 1/2 $, $ n = 1/2 $, $ a_0=1/4$, $ t_s = 1000$, with $ \eta_{max}= 7.90569 \pi $.}
    \label{plot3}
  \end{minipage}

  \label{fig:all_plots}
\end{figure}

The plots in Figs. \ref{plot1}, \ref{plot2}, and \ref{plot3} depict the Penrose diagrams for different parameters $a_0$ and $t_s$. The Misner-Sharp horizon is located at the edge of the gray region, and its behavior strongly depends on $a_0$ and $t_s$. In all cases, the trapping horizon initiates at $\eta = 0$ and $r = 0$. In Fig. \ref{plot1}, the trapping horizon appears to be spacelike, in Fig. \ref{plot2}, it is null while in Fig. \ref{plot3}, it is timelike \cite{Harada}.

Another analytical solution for the model given by (\ref{ScaleFactor4}) can be obtained for $ m = 1/2 $ and $ n= 3/2 $. The conformal time is:
\bea
\label{ConfTimeS2}
\eta = \frac{2 \sqrt{t}}{a_0 \sqrt{1 - \frac{t}{t_s}}}~,
\eea
and the range of $ \eta $ is the following:
\bea
\label{ConfTimeRangeS2}
0 \leq \eta \leq \infty~.
\eea
In this case, we also encounter a Big Bang singularity as the initial singularity for both the matter and gravitational metrics. As $t \rightarrow 0$, which corresponds to $\eta \rightarrow 0$, the scale factor $ a  \rightarrow 0 $, the scalar field $ \alpha \rightarrow 1 $ and both the density and the pressure become singular ($\rho \rightarrow \infty$, $p \rightarrow \infty$). When approaching the singularity at $t \rightarrow t_s$, the scale factor for matter, $a\rightarrow a_0 \sqrt{t_s}$, the scalar field $\alpha\rightarrow 0$, the density $\rho \rightarrow 0$, and the pressure module $\left| p \right| \rightarrow \infty$. Similarly, as time approaches the singularity, $t \rightarrow t_s$, the conformal time, $\eta$ becomes infinite. This singularity shares similarities with the Sudden Future Singularity (SFS), where $a(t_s) =$ const., $\rho =$ const., and the pressure $p \rightarrow \infty$.

The relation (\ref{ConfTimeS2}) can be inverted into
\bea
\label{InvertConfTimeS2}
t \left( \eta \right) = \frac{t_s {a_0}^2 {\eta}^2}{4 t_s + {a_0}^2 {\eta}^2}~.
\eea
%

The affine parameter (\ref{AffineParameter}) is:
\bea
\label{AffineParameterS2}
\lambda \left( t \right) = \frac{a_0}{24 t_s} \left[ \sqrt{t - \frac{t^2}{t_s}} \left( 14 tt_s -8 t^2 -3 {t_s}^2 \right) \right. + 3 \left. {t_s}^{\frac{5}{2}} \arcsin \left( \sqrt{\frac{t}{t_s}} \right)   \right]~.
\eea
Thus, at the Big Bang singularity, $\lambda(0) = 0$, and at the singularity time $t_s$, $\lambda(t_s) = (1/16) a_0 \pi {t_s}^{3/2}$. In this case the Misner-Sharp mass $M$ results from the following expression:
\bea
\label{SharpMisnerMassS2}
\frac{2M}{a \left( t \right) \alpha \left( t \right)} = \frac{ {a_0}^2 r^2 \sqrt{1 - \frac{t}{t_s}} {\left( t_s - 4 t \right)}^2 {\left( t_s - t \right)}^5}{4 t {t_s}^7} ~,
\eea
and the trapping past/future regions occur for
\bea
r > \frac{2 \sqrt{t} {t_s}^{\frac{7}{2}}}{ a_0 {\left( 1 - \frac{t}{t_s} \right)}^{\frac{1}{4}} \left( t_s - 4 t \right) { \left( t_s - t \right)}^{\frac{5}{2}}}~.
\eea

\begin{figure}

  \begin{minipage}[b]{0.37\textwidth}
    \includegraphics[width=\linewidth]{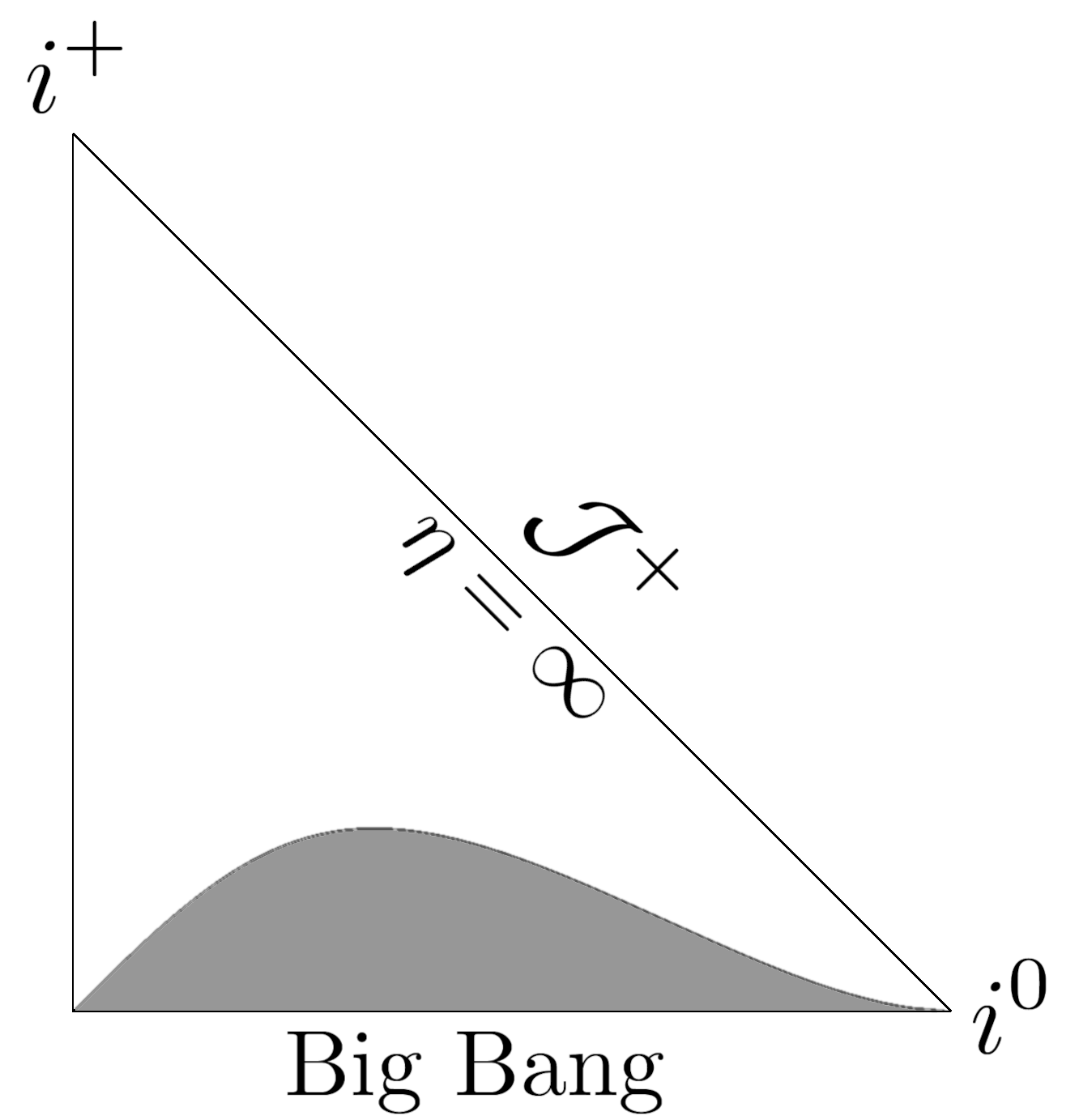}
    \captionsetup{width=1.3\linewidth} 
    \caption{Penrose diagram for the model given by  (\ref{ScaleFactor4}) with $ m = 1/2 $, $ n = 3/2 $, $ a_0=1$, $ t_s = 1$.}
    \label{plot4}
  \end{minipage}
  \hspace{0.1\textwidth}
  \begin{minipage}[b]{0.37\textwidth}
    \includegraphics[width=\linewidth]{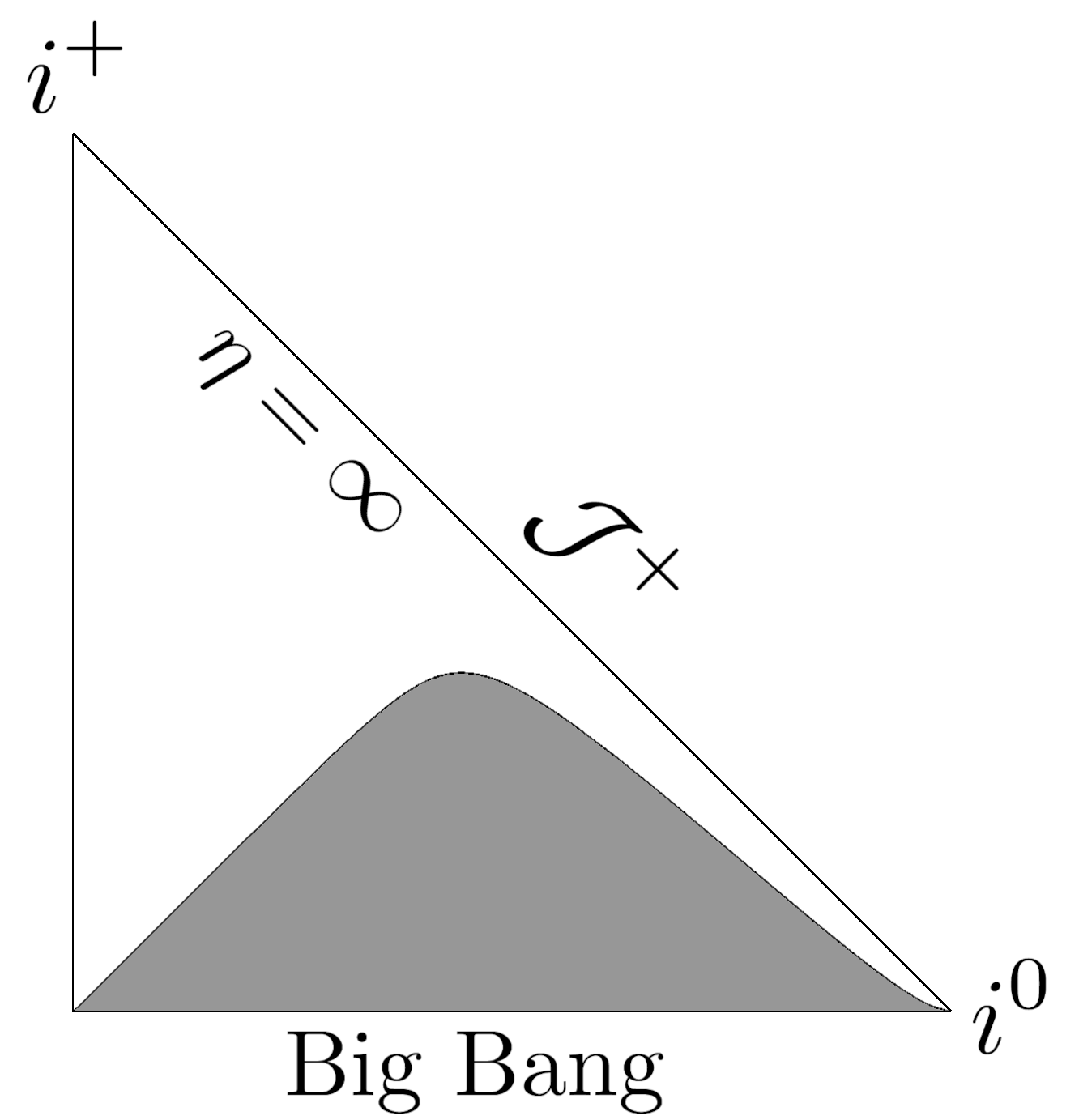}
     \captionsetup{width=1.3\linewidth}
    \caption{Penrose diagram for the model given by  (\ref{ScaleFactor4})  with $ m = 1/2 $, $ n = 3/2 $, $ a_0=1$, $ t_s = 100$.}
    \label{plot5}
  \end{minipage}
  \hspace{0.1\textwidth}
  \begin{minipage}[b]{0.37\textwidth}
    \includegraphics[width=\linewidth]{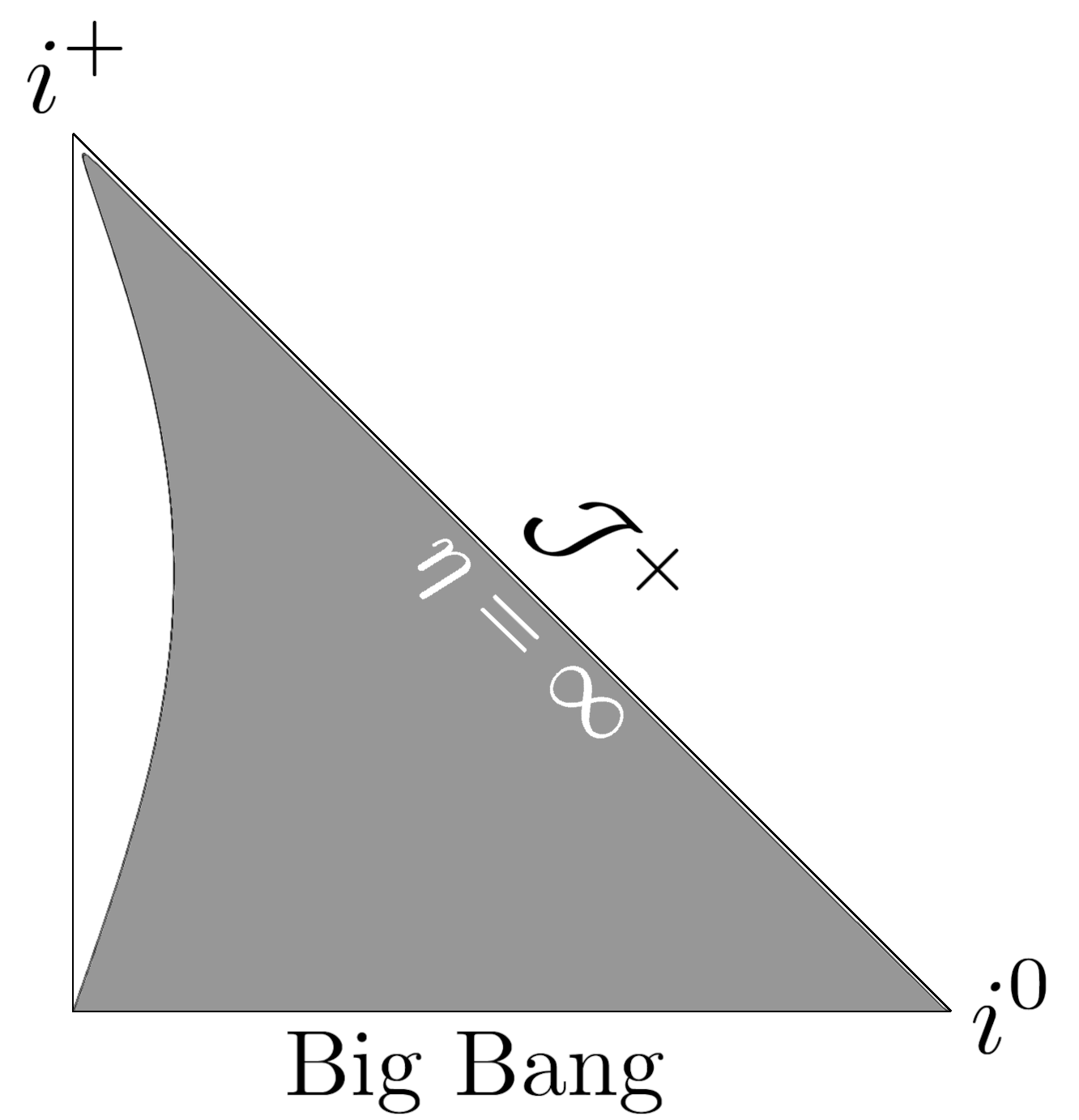}
     \captionsetup{width=1.3\linewidth}
    \caption{Penrose diagram for the model given by  (\ref{ScaleFactor4})  with $ m = 1/2 $, $ n = 3/2 $, $ a_0=1/8$, $ t_s = 100$.}
    \label{plot6}
  \end{minipage}
  \hspace{0.2\textwidth}
  \begin{minipage}[b]{0.37\textwidth}
    \includegraphics[width=\linewidth]{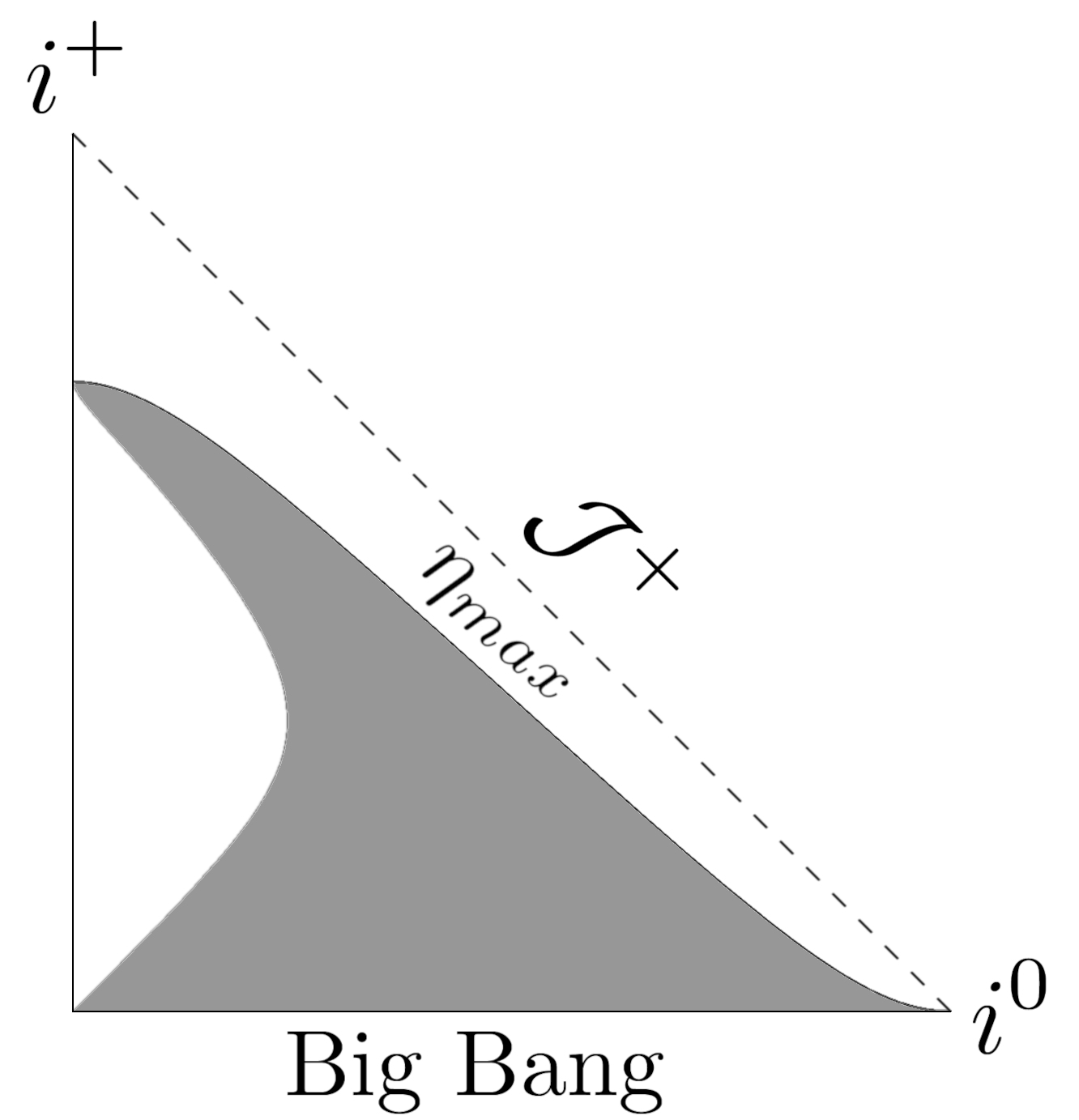}
     \captionsetup{width=1.3\linewidth}
    \caption{Penrose diagram for the model given by (\ref{ScaleFactor4}) with $ m = 1/2 $, $ n = -1 $, $ a_0=2$, $ t_s = 10$.}
    \label{plot7}
  \end{minipage}
  \label{fig:all_plots}
\end{figure}

The plots in Figs. \ref{plot4}, \ref{plot5}, and \ref{plot6} depict the Penrose diagrams for $ m = 1/2 $ and $ n= 3/2 $. The Misner-Sharp horizon is located at the edge of the gray region. In Figs. \ref{plot3} and \ref{plot6}, for $a_0 < 1$ and $t_s \gg a_0$, the horizon is truncated in the vicinity of $\mathcal{J}^+$. As $t_s \rightarrow \infty$, the shape of the trapping horizon on the left side of the plot persists, extending towards $i^+$ without interrupting the region near $\mathcal{J}^+$ and exhibiting timelike behavior. For $a_0 = 1$ and $t_s \gg a_0$, the trapping horizon in Figs. \ref{plot2} and \ref{plot5} approaches null as $t_s \rightarrow \infty$. For $ a_0 > 1 $, the spacelike behavior prevails and becomes less dependent on the $ t_s $ parameter.
For $ n = 1/2 $ the nature of the singularity induced by varying constants is similar to typical exotic singularitiy. Since $\eta_{max}=$ const., the singularity emerges similarly to standard exotic singularities \cite{Dabrowski2}. In both cases ($ n = 1/2 $ and $ n = 3/2 $), the dynamical gravitational constant $G$ (\ref{gvarying}) and the speed of light $c$ (\ref{cvarying}) approach infinity as the singularity is reached.

The last analytical solution we found is for $m = 1/2$ and $n = -1$. The conformal time (\ref{conftime}) in this model is as follows:
\bea
\label{ConfTimeS3}
\eta = \frac{2 \sqrt{t} \left( 3 t_s - t \right)}{3 a_0 t_s}~.
\eea
The scenario starts with the Big Bang at $t \rightarrow 0$, which corresponds to $\eta \rightarrow 0$, and the evolution of the Universe ends for $t \rightarrow t_s$, representing in conformal time notation $\eta = 4 \sqrt{t_s} / (3 a_0)$. The inverted form of relation (\ref{ConfTimeS3}) is expressed as:
\bea
\label{InvertConfTimeS3}
t \left( \eta \right) = 2 t_s - \frac{\left( 1 + i \sqrt{3} \right) {t_s}^2}{{ X}^{\frac{1}{3}}} - \frac{1}{4} \left( 1 - i \sqrt{3} \right) {X^{\frac{1}{3}}}~,
\eea
where
\bea
X = 9 {a_0}^2 {t_s}^2 {\eta}^2 - 8 {t_s}^3 + 3 \sqrt{9 {a_0}^4 {t_s}^4 {\eta}^4 - 16 {a_0}^2 {t_s}^5 {\eta}^2 }~.
\eea

The affine parameter (\ref{AffineParameter}) in this case is the following:
\bea
\label{AffineParameterS3}
\lambda  a_0 {t_s}^{\frac{3}{2}} {arc \tanh} \left( \sqrt{ \frac{t}{t_s}} \right) - 2 a_0 \sqrt{t} {t_s}^2~,
\eea
so it changes from $ \lambda = 0 $ at the Big Bang to infinity at the future singularity, indicating geodesic incompleteness. The Misner-Sharp mass $M$ can be obtained from the following condition:
\bea
\label{SharpMisnerMassS3}
\frac{2M}{a \left( t \right) \alpha \left( t \right)} = - \frac{{a_0}^2 r^2 {t_s}^5 {\left( t + t_s \right)}^2}{ 4 t {\left( t_s - t \right)}^7}~,
\eea
which gives the trapping future/past region as
\bea
r > \frac{2 \sqrt{t} {\left( t_s - t \right)}^{\frac{7}{2}}}{ a_0 {t_s}^{\frac{5}{2}} \left( t + t_s \right)}~.
\eea

The Misner-Sharp horizon is on the left side of the grey region in Fig. \ref{plot7} and it is timelike.  The model initiates with Big Bang singularity and ends as $ t \rightarrow t_s $, where $ a \left( t_s \right) =$const., $ \alpha  \rightarrow \infty $, the density $ \rho \rightarrow \infty $ and the pressure module $ \left| p \right| \rightarrow \infty $. The future singularity displays behavior resembling the finite scale factor singularity (FSF) in the matter. Let us note that the models with $ n = 1/2 $ and $ n = -1 $ give qualitatively the same future singularity for the matter sector but the behavior of $ \alpha $ in both cases is different.
In case of the models with $ n > 0 $, as $t \rightarrow t_s$, the scalar field $ \alpha \rightarrow 0 $, resulting in the effective gravitational constant $ G $ and the speed of gravitational interactions propagation $c$ approaching infinity. On the other hand, for $ n < 0 $, as $t \rightarrow t_s$, $ \alpha  \rightarrow \infty $ which imply $ G \rightarrow 0 $ and $ c   \rightarrow 0 $.

\section{Conclusions}
\label{sec5}

In this paper we employed a specific type of bimetric model with varying constants to explore the conformal structure of spacetimes using Penrose diagrams and to study singularities. We focused on the gravitational metric which describes the causal structure for the gravitational field. Our results indicate that for $n = 1/2$, corresponding to a strong singularity according to  Kr\'olak and Tipler's definition, the nature of this singularity resembles that of the finite scale factor singularity (FSF). The Penrose diagrams exhibit similarities, although the Misner-Sharp horizon exhibits strong dependence on the initial parameters $a_0$ and $t_s$. We demonstrated that the behavior of the horizon depends on initial conditions and can be classified into three primary types. The nature of the Misner-Sharp horizon is almost identical for both $n = 1/2$ and $n = 3/2$. It is particularly intriguing that the result for $n = 3/2$, which qualifies as a strong singularity according to  Kr\'olak and Tipler's definition, resembles the sudden future singularity (SFS), typically considered weak. Although parameters such as scale factor, density, and pressure align with SFS, the nature of the singularity differs. The singularity manifests at $\mathcal{J}^+$, as opposed to $\eta =$ const. in SFS. Moreover, the Misner-Sharp horizon differs from that in the SFS case. The conformal structure for $m=1/2$ and $n=-1$ is identical to that of the standard exotic singularity for the matter. In both cases the singularity emerges on hypersurface which is spacelike and the behaviour of Misner-Sharp horizon is timelike.

Our bimetric model assumes $ \alpha \left( t \right) $ which can be interpreted as a scalar field. The dynamical gravitational constant and the dynamical speed of light in the gravitational metric depend on $ \alpha $, preventing a clear separation between the effects of the dynamical gravitational constant and the dynamical speed of light.\\

\noindent \textbf{Data Availability Statement:} All data generated or analysed during this study are included in this published article. \\
\textbf{Acknowledgments} We wish to thank Mariusz D\c{a}browski for discussions.

\end{document}